\documentclass[11pt]{article}

\usepackage{epsfig}
\usepackage{amsfonts,amssymb}

\renewcommand{\theequation}{\arabic{section}.\arabic{equation}}

\font\mybb=msbm10 at 12pt 
\def\bb#1{\hbox{\mybb#1}}

\def\bE {\bb{E}} 
 
\def\bM {\bb{M}}

\newcommand{\A}{\ensuremath{{\cal A}}}

\newcommand{\del}{\ensuremath{\partial}}

\newcommand{\half}{\ensuremath{\frac{1}{2}}}

\newcommand{\be}{\begin{equation}}
\newcommand{\ee}{\end{equation}}

\newcommand{\ba}{\begin{eqnarray}}
\newcommand{\ea}{\end{eqnarray}}

\newcommand{\ns}{\normalsize}

\begin{document}


\begin{titlepage}

\title{
   \hfill{\ns hep-th/0407156\\}
   \vskip 2cm
   {\Large\bf Type IIA Killing spinors and calibrations}
\\[0.5cm]}
   \setcounter{footnote}{0}
\author{
{\ns\large 
  \setcounter{footnote}{3}
  P.~M.~Saffin\footnote{email: P.M.Saffin@sussex.ac.uk}}
\\[0.5cm]
   {\it\ns Department of Physics and Astronomy, University of Sussex}
   \\
   {\ns Falmer, Brighton BN1 9QJ, UK} \\[0.2em] }
\date{}

\maketitle

\begin{abstract}\noindent
We consider the dimensional reduction of eleven dimensional supergravity
to type IIA in ten dimensions, and study the conditions for supersymmetry
in terms of $p$-form spinor bi-linears of the supersymmetry parameter.
For a bosonic solution to be supersymmetric these $p$-forms must satisfy
a set of differential relations, which we derive in full; the supersymmetry
variations of the dilatino give a set of algebraic relations which are
also derived. These results are then used to provide the generalized
calibration conditions for some of the basic brane solutions, we also
follow up a suggestion of and Hackett-Jones and Smith and present a calibration
condition for IIA supertubes. We find that a probe supertube satisfies this
bound but does not saturate it, with the bound successfully accounting for
the D0 charge of the supertube but not the string charge; we speculate that
there should be a stronger calibration inequality than the one given.
\end{abstract}

\thispagestyle{empty}

\end{titlepage}

\section{Introduction}
The utility of bi-linears formed from Killing spinors has been known for some
time, with them being used to present Bogomol'nyi bounds in supergravity
\cite{gibbons:82}, derive the full set of supersymmetric solutions to
supergravities \cite{tod:83} and to give constraints on the manifolds used in dimensional
reduction \cite{candelas:84}.
More recently this approach has been used to generate the
supersymmetric solutions to higher dimensional supergravities
\cite{Gauntlett:2002nw}\cite{Gauntlett:2003fk}\cite{Gutowski:2003rg}\cite{Caldarelli:2003pb},
by solving the differential equations of the $p$-form spinor bi-linears
rather than directly solving the Killing spinor equation; here we present the
differential and algebraic equations for the spinor bi-linears of IIA supergravity.
Theses $p$-forms also naturally lead one to the notion of $G$-structures, as
the existence of globally defined $p$-forms reduces the frame bundle SO(1,D-1) to some
$G$-sub-bundle, this can then be used 
to give classifications of supergravity solutions
\cite{gurrieri:0211}-\cite{Lukas:2004ip}. 

Calibrations \cite{Gibbons:1998hm} form an important 
part of string theory technology and our understanding has developed from
the basic calibrations, which describe minimal volume submanifolds, to generalized
calibrations \cite{Gutowski:1999tu} relevant to branes moving in background fluxes.
In \cite{Hackett-Jones:2003vz} it was shown how the differential relations of
the $p$-form bi-linears in eleven dimensional supergravity could be combined in
a natural way with the flat space supersymmetry algebra to give a proposal for the
calibration bound of M2 and M5 branes. Applying these ideas to IIB supergravity
in ten dimensions \cite{Hackett-Jones:2004yi} lead to a proposal for the 
calibration bound of giant gravitons, the first example of a calibration for non-static
branes. As an application of our results we consider a case suggested by
\cite{Hackett-Jones:2004yi} as another example of a solution which could lead to
a calibration condition for non-static brane, the supertube \cite{Mateos:2001qs},
thought of as a configuration of D0 branes and parallel strings blown up to
a cylindrical D2 brane.
We propose such a bound but believe that it is not the full story as 
the bound is satisfied by a probe supertube but not saturated; the bound correctly
captures the D0 charge but misses the string charge.

The outline of the paper is as follows. Section \ref{MtoIIA} gives the $p$-form
spinor bi-linears that are allowed in IIA supergravity and relates them to those
from eleven dimensions. The differential relations for these $p$-forms are then derived
from the analogous 11D relations in section \ref{diffRelations}. The algebraic
constraints which follow from the IIA Killing spinor equation are derived
in section \ref{algRelations}. The final section describes how calibration conditions
are derived from the differential relations and the supersymmetry algebra, we
show some simple examples of how this works and end with the supertube example.
In the appendices we give our conventions in deriving IIA from eleven dimensions and
also provide some of the basic IIA solutions which act as a check for our
equations.

\section{IIA from eleven dimensions}
\label{MtoIIA}

It is well known, \cite{Campbell:zc}\cite{Huq:1983im}\cite{Giani:wc}
that IIA supergravity can be derived by dimensional reduction
from the eleven dimensional supergravity of Cremmer-Julia-Scherk (CJS)
\cite{Cremmer:1978km}. 
This allows us to use the results of \cite{Gauntlett:2002fz} to obtain the set
of bi-linears in IIA.
In order for our conventions to conform to that
of \cite{Gauntlett:2002fz} we show in the appendix both supergravities with
their corresponding supersymmetry transformations.

When there is a single Killing spinor one can use the symmetry properties of
the Dirac matrices
in eleven dimensions to define the following spinor bi-linears (objects with a hat
are eleven dimensional),
\ba
\hat{K}     &=&\bar{\hat\epsilon}\hat\Gamma_{\hat{a}}\hat\epsilon \;E^{\hat a},\\
\hat{\Omega}&=&\frac{1}{2}\bar{\hat\epsilon}\hat\Gamma_{\hat{a_1}\hat{a_2}}\hat\epsilon \;E^{\hat{a_1}\hat{a_2}},\\
\hat{\Sigma}&=&\frac{1}{5!}\bar{\hat\epsilon}\hat\Gamma_{\hat{a_1}...\hat{a_5}}\hat\epsilon \;E^{\hat{a_1}...\hat{a_5}}.
\ea
This is the full set, with the other rank $p$-forms either vanishing or being related by
Hodge duality.

To reduce these to ten dimensions we note from appendix \ref{AppA} that the 11D and 10D
vielbein are related by $E^a=\exp(-\phi/3)e^a$. This takes us to the string frame, which is more natural
in this context as the supersymmetry transformations (\ref{IIAdilVar}), (\ref{IIAgravVar})
have some degree of uniformity amongst the field strengths. With this we find
\ba
\label{11DbilinearsK}
\hat{K}     &=&\exp(-2\phi/3)K+\exp(-\phi/3)XE^{\underline{z}},\\
\hat{\Omega}&=&\exp(-\phi)\Omega+\exp(-2\phi/3)\tilde{K}\wedge E^{\underline{z}},\\
\label{11DbilinearsSigma}
\hat{\Sigma}&=&\exp(-2\phi)\Sigma+\exp(-5\phi/3)Z\wedge E^{\underline{z}},
\ea
where we have introduced the following 10D bi-linears
\ba
\label{10DbilinearsX}
X&=&\bar{\epsilon}\Gamma_{\underline{z}}\epsilon,\\
\label{10DbilinearsK}
K&=&\bar{\epsilon}\Gamma_{a}\epsilon \;e^{a},\\
\label{10DbilinearsKtilde}
\tilde{K}&=&\bar{\epsilon}\Gamma_{a}\Gamma_{\underline{z}}\epsilon \;e^{a},\\
\Omega&=&\frac{1}{2}\bar{\epsilon}\Gamma_{a_1a_2}\epsilon \;e^{a_1a_2},\\
Z&=&\frac{1}{4!}\bar{\epsilon}\Gamma_{a_1...a_4}\Gamma_{\underline{z}}\epsilon \;e^{a_1...a_4},\\
\label{10DbilinearsSigma}
\Sigma&=&\frac{1}{5!}\bar{\epsilon}\Gamma_{a_1...a_5}\epsilon \;e^{a_1...a_5}.
\ea
In doing this we have chosen to represent the two Majorana-Weyl spinor parameters, $\epsilon^\pm$, of
IIA by the single Majorana spinor $\epsilon=\epsilon^+ +\epsilon^-$. An alternative set of
bi-linears could have been defined in terms of the Majorana-Weyl components,
$K^{++}=\bar{\epsilon^+}\Gamma_{a}\epsilon^+ \;e^{a}$,... but these can be constructed from
linear combinations of (\ref{10DbilinearsX}-\ref{10DbilinearsSigma}) and so are equivalent. We shall find
the former set more useful
as $\epsilon$ descends directly from $\hat\epsilon$ allowing us to use the relations 
of \cite{Gauntlett:2002fz} to derive differential equations for (\ref{10DbilinearsX}-\ref{10DbilinearsSigma}),
which we now do.

\section{Differential relations.}
\label{diffRelations}

Now we have the set of $p$-forms we can derive the differential relations they must satisfy
if the solution is to be supersymmetric. As described in \cite{Gauntlett:2002fz} 
this is achieved by using the vanishing of the susy variation (\ref{11Dsusy}).
One finds that the 11D bi-linears must solve
\ba
d\hat{K}&=&\frac{2}{3}\hat{\Omega}\hat\lrcorner G+\frac{1}{3}\hat{\Sigma}\hat\lrcorner\hat{\star} \hat G,\\
d\hat{\Omega}&=&\hat{K}\hat\lrcorner \hat G,\\
d\hat{\Sigma}&=&\hat{K}\hat\lrcorner \hat\star G-\hat{\Omega}\wedge \hat G.
\ea
From these we can derive the analogous equations for the IIA $p$-forms, (\ref{10DbilinearsX}-\ref{10DbilinearsSigma}).
We find the following,
\ba
\label{IIAdiff1}
\exp(-\phi/3)d\left[\exp(\phi/3)X\right]&=&\frac{2}{3}\Omega\lrcorner H
                                           +\frac{1}{3}\exp(\phi)\Sigma\lrcorner\star\tilde{G},\\
\label{IIAdiff2}
\exp(2\phi/3)d\left[\exp(-2\phi/3)K\right]&=& \frac{2}{3}\tilde{K}\lrcorner H+\frac{1}{3}\Sigma\lrcorner\star H
                                             +\exp(\phi)\left[
                                                  \frac{2}{3}\Omega\lrcorner\tilde{G}-XF
                                                  +\frac{1}{3}Z\lrcorner\star\tilde{G}\right],\\
\label{IIAdiff3}
d\tilde{K}&=&K\lrcorner H,\\
\label{IIAdiff4}
\exp(\phi)d\left[\exp(-\phi)\Omega\right]&=&-XH+\exp(\phi)\left[\tilde{K}\wedge F
                                               +K\lrcorner\tilde{G}\right],\\
\label{IIAdiff5}
\exp(\phi)d\left[\exp(-\phi)Z\right]&=&-\Omega\wedge H+\exp(\phi)\left[6K\lrcorner\star\tilde{G}
                                                                       -\tilde{K}\wedge\tilde{G}\right],\\
\label{IIAdiff6}
\exp(2\phi)d\left[\exp(-2\phi)\Sigma\right]&=&K\lrcorner\star H
                             +\exp(\phi)\left[-Z\wedge dA+X\star\tilde{G}-\Omega\wedge\tilde{G}\right],
\ea
where the various field strengths are defined in appendix \ref{AppA}. An alternative route to these
equations is to directly consider the derivative of the forms in (\ref{10DbilinearsX}-\ref{10DbilinearsSigma})
and use the vanishing of 10D gravitino variation 
(\ref{IIAgravVar}) to replace the $\nabla_m\epsilon$ terms.

An important result that also comes from the analysis of $\nabla_a K_b$ is that
\ba
\nabla_{(a}K_{b)}&=&\frac{1}{3}\eta_{ab}K\lrcorner d\phi=0,
\ea
where the last equality follows from (\ref{IIAalg1}) to be derived in the next section. This
tells us then that one of the vector bi-linears, $K$, is in fact a Killing vector.
That such a spinor bi-linear is Killing is also true in eleven dimensions \cite{Gauntlett:2002fz}
and in IIB theory \cite{Hackett-Jones:2004yi}, this has important consequences when it comes
to constructing the calibration forms.

\section{Algebraic relations.}
\label{algRelations}

Whilst it was possible to simply translate the differential relations of \cite{Gauntlett:2002fz} into
differential relations relevant to the IIA $p$-forms we also have a set of algebraic constraints
coming from the vanishing of the susy
variation of the dilatino, (\ref{IIAdilVar}). To derive these we take (\ref{IIAdilVar}) and act
on the left with $\bar\epsilon$, $\bar\epsilon\Gamma^i$, $\bar\epsilon\Gamma^{ij}$, $\bar\epsilon\Gamma^{ijk}$
and $\bar\epsilon\Gamma^{ijkl}$ to give
\ba
\label{IIAalg1}
0&=&K\lrcorner d\phi,\\
\label{IIAalg2}
0&=&d\phi\lrcorner\Omega-\half H\lrcorner Z
     +\frac{1}{4}\exp(\phi)\left[3\tilde{K}\lrcorner F+\tilde{G}\lrcorner\Sigma\right],\\
\label{IIAalg3}
0&=&-d\phi\wedge K+\half H\lrcorner\star\Sigma+\half\tilde{K}\lrcorner H
  +\frac{1}{4}\exp(\phi)\left[-3F\lrcorner Z+3XF+\tilde{G}\lrcorner \star Z-\Omega\lrcorner\tilde{G}\right],\\
\label{IIAalg4}
0&=&(d\phi\wedge\Omega)^{ijk}-\frac{3}{4}H_{bc}^{\;\;\;[i}Z^{jk]bc}+\half XH^{ijk}\\\nonumber
 &~&+\frac{1}{4}\exp(\phi)\left[ 3(F\lrcorner\star\Sigma)^{ijk}+3(F\wedge\tilde{K})^{ijk}
                            -\half\tilde{G}_{bcd}^{\;\;\;\;\;\;[i}\Sigma^{jk]bcd}
                            +(K\lrcorner\tilde{G})^{ijk}\right],\\
\label{IIAalg5}
0&=&(d\phi\lrcorner\Sigma)^{ijkl}-H_{bc}^{\;\;\;\;[i}\star\Sigma^{jkl]bc}-\half(H\wedge\tilde{K})^{ijkl}\\\nonumber
   &~&+\exp(\phi)\left[-3F_b^{\;\;[i}Z^{jkl]b}+\frac{1}{6}\tilde{G}_{bcd}^{\;\;\;\;\;\;[i}\star Z^{jkl]bcd}
                       -\tilde{G}_d^{\;\;[ijk}\Omega^{l]d}\right].
\ea
These are the full set of relations which can be derived from (\ref{IIAdilVar});
if one hits (\ref{IIAdilVar}) with more the four $\Gamma$ matrices the resulting
relations will be dual to the above equations.

Although we can get no more algebraic relations from (\ref{IIAdilVar}) the fact that our
$p$-forms are bi-linear implies they must have certain relations amongst each other.
These can be constructed using Fiertz identities and they hold irrespective of supersymmetry.
Although we will not present anything like an exhaustive, list we give here two such examples
to illustrate the point.
From \cite{Gauntlett:2002fz} we have
\ba
K\hat\lrcorner\hat\Omega&=&0,\\
\hat{K}\hat\lrcorner\hat\Sigma&=&\half\hat\Omega\wedge\hat\Omega,
\ea
which we can convert into IIA language to get
\ba
K\lrcorner\tilde{K}&=&0,\\
K\lrcorner\Omega-X\tilde{K}&=&0,
\ea
and
\ba
XZ+K\lrcorner\Sigma&=&\half\Omega\wedge\Omega,\\
K\lrcorner Z&=&\Omega\wedge\tilde{K}.
\ea
respectively.

We have now given the full set of relations, differential and algebraic, which must be satisfied
by a supersymmetric solution of IIA supergravity. As a check that we have arrived at the
correct set of equations appendix \ref{AppB} provides some of the basic IIA solutions,
giving the spinor bi-linears. 

\section{Calibration conditions.}
\subsection{strings}
Following \cite{Hackett-Jones:2003vz} we find out how to derive calibration conditions
for some of the branes in IIA supergravity. We shall start with the simplest, namely the
F1 string. The super-Poincar\'e algebra in eleven flat dimensions,
with a probe M2 brane, can be dimensionally
reduced to give the algebra in ten flat dimensions with a probe string,
\ba
\label{F1algebra}
\{ Q_\alpha,Q_\beta \}&=&(C\Gamma^\mu)_{\alpha\beta}P_\mu
                \pm(C\Gamma_\mu\Gamma_{\underline{z}})_{\alpha\beta} Z^\mu,
\ea
where
\ba
Z^\mu=\int dX^\mu,
\ea
and the integration is over the spatial direction of the string.
If we now multiply (\ref{F1algebra}) by $\epsilon_0^\alpha\epsilon_0^\beta$,
for constant $\epsilon_0$, we have that
\ba
\label{flatF1calib}
(Q\epsilon_0)^2&=&K_0^\mu P_\mu\pm\tilde{K}_{0\mu}\int dX^\mu
               =\int d\sigma K_0^\mu p_\mu \pm \int \tilde K_0.
\ea
Where we have introduced a momentum density, $p_\mu$,
and the notation $K_0$ and $\tilde K_0$ come from (\ref{10DbilinearsK}),
(\ref{10DbilinearsKtilde}) applied to the constant spinor $\epsilon_0$,
$\sigma$ is the the spatial co-ordinate on the string world volume. By writing it in this
form we see how this equation should be generalized to curved space. We expect
the correction to the super-algebra to be a topological term reflecting the
charge of the probe, so the aim is to take the flat space expression $\int K_0$
and write it as the integral of a closed form. Now we use the fact that $K$
is a Killing vector and that the definition of the Lie derivative on forms,
\ba
{\cal L}_K\alpha&=&d(K\lrcorner\alpha)+K\lrcorner d\alpha.
\ea
From (\ref{IIAdiff3}) we have that $d(K\lrcorner H)$ and as $H=dB$ then
$K\lrcorner dH=0$ giving ${\cal L}_K H=0$, so we may choose a gauge in which
$B$ mirrors the symmetry of its field strength
in that ${\cal L}_K B=0$. We may therefore use this gauge and
rewrite (\ref{IIAdiff3}) as $d(\tilde K+K\lrcorner B)=0$,
which gives us the closed form we were looking for. The proposal, therefore, for
the curved version of (\ref{flatF1calib}) is 
\ba
\label{curvedF1calib}
(Q\epsilon)^2&=&\int d\sigma K.p \pm \int (\tilde K+K\lrcorner B)
\ea
where now we use the Killing spinor $\epsilon$. This then leads to the calibration bound
\ba
\label{stringBound}
\int d\sigma K.p \ge \mp \int (\tilde K+K\lrcorner B),
\ea
and a standard argument shows that a calibrated cycle (one for which
the bound is saturated) minimizes
$\int d\sigma K.p$ in its homology class.

As an example we could consider a string probe in the background of a stack
of strings whose solution is given by (\ref{F1section}). There we find that
\ba
\label{stringForm}
\tilde K+K\lrcorner B=dx,
\ea
with $dx$ being the spatial direction of the string, this is clearly closed.
With these relations we can check to see if a string probe in the background
of a multi-string solution (Appendix \ref{AppB}) saturates the calibration bound.
For this we identify $K.p$ with the Hamiltonian density of the probe, which we
can calculate from the string action
\ba
S_{F1}&=&-\int d^2\sigma\sqrt{-\gamma}-\int {\cal P}[B].
\ea
Where $\gamma_{\mu\nu}$ is the induced metric on the world-volume of the probe and
${\cal P}[B]$ is the pull back of the spacetime field, $B$, to the string world-volume.
We shall look at a probe string oriented in the same way as the background strings,
the $t-x$ plane as in (\ref{stringSolution}). For the world-volume co-ordinates
$(\tau,\sigma)$ we choose the natural gauge $\tau=t$, $\sigma=x$
which then leads to the energy density ${\cal H}=1=K.p$. Using (\ref{stringForm})
we see that the string probe saturates the calibration bound (\ref{stringBound}).

\subsection{D2-branes}

For D2-branes the situation is slightly more complicated as we have been unable
to find the general calibration condition
for any given background fields, but we can look at the bound in
any specific case. As an example we look at the bound for a probe D2 in the
background of a stack of gravitating D2-branes, (\ref{D2section}). In this case
we see that $H=0$, $F=0$ so (\ref{Gtilde}) gives $\tilde{G}=dC$ and 
(\ref{IIAdiff4}) shows ${\cal L}_K dC=d(K\lrcorner dC)+K\lrcorner ddC=0$.
Thus, $K$ represents a symmetry of the field strength $dC$, in which
case we may pick a gauge where $C$ also has this symmetry and choose ${\cal L}_K C=0$. 
In that gauge then we have
${\cal L}_K C=d(K\lrcorner C)+K\lrcorner dC=0$ and (\ref{IIAdiff4}) gives us 
the following closed 2-form,
\ba
d\left[\exp(-\phi)\Omega+K\lrcorner C\right]=0.
\ea
The calibration bound again comes from the supersymmetry algebra. The terms relevant
for D2-branes in the flat 10D algebra are.
\ba
\label{D2algebra}
\{ Q_\alpha,Q_\beta \}&=&(C\Gamma^\mu)_{\alpha\beta}P_\mu
                \pm(C\Gamma_{\mu\nu})_{\alpha\beta} Z^{\mu\nu},
\ea
where
\ba
Z^{\mu\nu}=\int dX^\mu \wedge dX^\nu.
\ea
Going through the same procedure as the probe string we are led to suggest
the following calibration bound for a probe D2 in the background of a stack
of D2-branes,
\ba
\label{D2Bound}
\int d^2\sigma K.p \ge \mp \int (\exp(-\phi)\Omega+K\lrcorner C).
\ea
Note that the D2-brane stack given in \ref{D2section} has
\ba
\label{D2Form}
\exp(-\phi)\Omega+K\lrcorner C=dx^{12}+dy^{12}+dy^{34}+dy^{56},
\ea
where $dx^{12}$ is the spatial orientation of the stack and 
$d\underline {y}$ are transverse directions.
Again, this is clearly closed. As for the string, we can check to see if a D2
probe in the background of a stack of D2-branes, appendix \ref{AppB}, saturates
the calibration bound (\ref{D2Bound}). We shall orient our probe in the same
direction as the stack, $(t,x^1,x^2)$, and choose a gauge where the world-volume
co-ordinates match the spacetime ones, $(\tau=t,\sigma^1=x^1,\sigma^2=x^2)$.
Taking the action for a D2 brane
\ba
\label{D2action}
S_{D2}&=&-\int\exp(-\phi)\sqrt{-\gamma}-\int {\cal P}[C].
\ea
allows us to calculate the energy density to be
${\cal H}=1=K.p$. Now, using (\ref{D2Form}), we see that the
calibration bound (\ref{D2Bound}) is saturated. In fact, (\ref{D2action}) is not the
full world volume action of the D2-brane as there can be a Born-Infeld field, $F_{BI}$,
living on the world-volume which changes the action to
\ba
S&=&-\int\exp(-\phi)\sqrt{-det(\gamma+{\cal F})}+\int({\cal P}[C]+{\cal P}[A]\wedge{\cal F}),
\ea
where ${\cal F}$ is the gauge invariant world-volume field strength, ${\cal F}={\cal P}[B]+F_{BI}$.
This world-volume field strength must be reflected in the calibration condition, and as
$F_{BI}$ is already a closed two-form we would anticipate that (\ref{D2Bound}) should become
\ba
\label{D2Boundb}
\int d^2\sigma K.p \ge \mp \int (\exp(-\phi)\Omega+K\lrcorner C+F_{BI}).
\ea
This type of term has already been seen in the M5 brane calibration conditions,
\cite{Barwald:1999ux}\cite{Hackett-Jones:2003vz}, and we shall see evidence in the next section
that it should be there.

\subsection{supertubes}
Now we come to a more substantial example, following a suggestion in \cite{Hackett-Jones:2004yi},
that of a supertube \cite{Mateos:2001qs}.
This is a nice case for a number of reasons: it is sufficiently complex so as
to excite all the field strengths, thereby proving a good check on our relations;
from the point of view of the D2-brane there are world volume fields turned on,
giving a check to the Born-Infeld term which did not follow from the 
flat space susy algebra;
it is also not a static solution, unlike most other calibrated branes - except
for the giant graviton \cite{Hackett-Jones:2004yi}.

The supergravity version of supertubes was presented in \cite{Emparan:2001ux} and is
obtained by dimensionally reducing a solution found in \cite{Gauntlett:1998kc},
which describes the intersection of two rotating M5 branes and an M2 brane.
We have given the IIA solution in appendix \ref{supertubesection}.

As the supertube is to be considered as a D2 brane we shall be looking for a closed
two-form living in the solution, composed from spinor bi-linears and the fields which
are excited.
The first thing to note is that while neither $d(K\lrcorner C)$ nor $K\lrcorner dC$ vanish, their sum does,
giving ${\cal L}_K dC=0$. As usual then, we may choose a gauge where 
$C$ matches the symmetry of its field strength, ${\cal L}_K C=0$.
For the supertube we have from appendix \ref{supertubesection}
that $K\lrcorner A-\exp(-\phi)=-1$, using this and 
(\ref{IIAdiff3}) shows that (\ref{IIAdiff4}) leads to
\ba
d\left[\exp(-\phi)\Omega+K\lrcorner C+\tilde{K}\wedge A+B\right]&=&0,
\ea
with the supertube solution giving
\ba
\label{supertubeForm}
\exp(-\phi)\Omega+K\lrcorner C+\tilde{K}\wedge A+B&=&-dt\wedge dx,
\ea
This therefore suggests that the calibration bound in this background becomes
\ba
\int d^2\sigma K.p \ge \mp \int \left[\exp(-\phi)\Omega+K\lrcorner C+\tilde{K}\wedge A+B\right].
\ea
However, this has not taken into account the world-volume Born-Infeld field on the D2
and so, in analogy with the M5 calibration bound \cite{Barwald:1999ux}\cite{Hackett-Jones:2003vz}
we put forward the following bound
\ba
\label{supertubeBound}
\int d^2\sigma K.p \ge \mp \int \left[\exp(-\phi)\Omega+K\lrcorner C+\tilde{K}\wedge A+B+F_{BI}\right],
\ea
with $F_{BI}$ being the Born-Infeld field strength. We are now in a position to test this bound by
placing a probe supertube in the background. Following \cite{Emparan:2001ux} we 
choose our supertube to be cylindrical and write the metric on $d\underline{y}^2$ 
of (\ref{supertubeMetric}) as
\ba
d\underline{y}^2&=&dr^2+r^2d\varphi^2+d\rho^2+\rho^2 d\Omega_{(5)}^2.
\ea
The form of the harmonic functions is given in \cite{Emparan:2001ux}, they place
the supertube at $r=R$ and $\rho=0$ so it occupies the $t-x-\varphi$ direction,
the one form ${\cal A}$ is given by ${\cal A}=\tilde{\cal A}(r,\rho)d\varphi$.
Now consider a probe
D2 with the following Born-Infeld field strength
\ba
\label{BIansatz}
F_{BI}&=&E_{BI}dt\wedge dx+B_{BI}dx\wedge d\varphi.
\ea
The gauge invariant world volume field strength is
\ba
{\cal F}&=&F_{BI}+{\cal P}[B]:={\cal E}dt\wedge dx+{\cal B}dx\wedge d\varphi.
\ea
Now we may take the probe action
\ba
S&=&-\int\exp(-\phi)\sqrt{-det(\gamma+{\cal F})}+\int({\cal P}[C]+{\cal P}[A]\wedge{\cal F}),
\ea
to find the following Lagrange density (in the usual physical gauge) \cite{Emparan:2001ux}
\ba
{\cal L}&=&-U^{\frac{1}{2}}V^{-1}\sqrt{Vr^2(U^{-2}-{\cal E}^2)+U^{-1}({\cal B}-\tilde{\cal A}{\cal E})^2}
           +V^{-1}({\cal B}-\tilde{\cal A}{\cal E})-B_{BI}.
\ea
To find the Hamiltonian we first need the conjugate of $E_{BI}$, ${\cal D}$.
\ba
{\cal D}&=&\frac{\del {\cal L}}{\del E_{BI}}
          =\frac{\del {\cal L}}{\del {\cal E}}
          =\frac{U^{\frac{1}{2}}r^2{\cal E}+U^{-\frac{1}{2}}V^{-1}({\cal B}-\tilde{\cal A}{\cal E})\tilde{\cal A}}
                {\sqrt{Vr^2(U^{-2}-{\cal E}^2)+U^{-1}({\cal B}-\tilde{\cal A}{\cal E})^2}}
           -V^{-1}\tilde{\cal A},
\ea
As a supertube has $E_{BI}=1$ \cite{Emparan:2001ux}, then we find that the Hamiltonian is
\ba
\label{Htube}
\int dxd\varphi{\cal H}&=&\int dxd\varphi\left[{\cal D}E-{\cal L}\right]=\int dxd\varphi\left[{\cal D}+B_{BI}\right],
\ea
with ${\cal D}=R^2/B_{BI}$.
In terms of the calibration condition (\ref{supertubeBound}) we identify (\ref{Htube}) 
with $\int d^2\sigma K.p$.
On the right hand side of the bound we use (\ref{supertubeForm}) and (\ref{BIansatz}) to get
\ba
\int \left[\exp(-\phi)\Omega+K\lrcorner C+\tilde{K}\wedge A+B+F_{BI}\right] = \int dxd\varphi B_{BI}.
\ea
So, while the bound is satisfied, it is not saturated. As noted in \cite{Emparan:2001ux} ${\cal D}$ corresponds
to the string charge of the probe and $B_{BI}$ the D0 brane charge, so we see that the bound has
successfully accounted for the D0 charge whilst missing the string charge. We believe that the 
supertube should saturate some calibration bound; if so, then there will be a stronger inequality than
presented in (\ref{supertubeBound}).

\section{Conclusion}

In this paper we have described how one can construct $p$-forms from the
Killing spinors of IIA supergravity using those of CJS supergravity in eleven
dimensions. 
As is to be expected from dimensional reduction there are more of these
bi-linears than in the parent theory, with one scalar, two vectors, a two-form,
a four-form and a five-form, along with their Hodge duals.
The set of differential relations satisfied by these spinor
bi-linears was derived and shown
to follow from the analogous relations in 11D, as given
in \cite{Gauntlett:2002fz}. 
Unlike the bi-linears of CJS supergravity, one of the killing spinor equations,
the variation of the dilatino, gave a set of algebraic constraints, in
concert with the algebraic constraints coming from Fiertz identities.
We found the full set of these dilatino constraints and gave some examples
of the Fiertz relations.

As an application of these results we considered a technique introduced in
\cite{Hackett-Jones:2003vz} for proposing calibration bounds. We used this
to give the bound for a string in a general background, but for the D2 brane
we could only find background-specific results. In particular we applied
our equations to the supertube of \cite{Emparan:2001ux} which lead to
a putative calibration bound. A probe supertube was found to satisfy the
bound, but not saturate it, leading to the suspicion that a stronger bound
should exist. This stemmed from a deficiency of the technique in that world
volume fields are not naturally accounted for.

\vspace{1cm}
\noindent
Note added

\noindent
After the completion of this work there appeared a pre-print by
Cascales and Uranga \cite{Cascales:2004qp} proposing another method for finding 
calibration bounds, we hope that their results when applied to IIA will
strengthen the supertube bound found here.

\vspace{1cm}
\noindent
{\large\bf Acknowledgements} We would like to thank Douglas Smith
for discussions. The author is supported by
a PPARC Advanced Fellowship.\\

\vskip 1cm
\appendix{\noindent\Large \bf Appendices}
\renewcommand{\theequation}{\Alph{section}.\arabic{equation}}
\setcounter{equation}{0}

\section{D=11, 10 supergravities and conventions.}
\label{AppA}

We follow the conventions of \cite{Gauntlett:2002fz} in using (-,+,+,...) as our
spacetime signature with the alternating symbol $\epsilon_{012...}=+1$. The inner
product of $q$-forms with $p(<q)$-forms is
\ba
(\alpha_p\lrcorner\beta_q)_{a_1...a_{q-p}}=(1/p!)\alpha^{b_1...b_p}\beta_{b_1...b_pa_1...a_{q-p}},
\ea
and the Hodge dual is defined by
\ba
\star\alpha_{a1...a_{D-p}}&=&(1/p!)\epsilon_{a1...a_{D-p}}^{\qquad\quad b_1...b_p}\alpha_{b_1...b_p}.
\ea
Flat indices are given by Roman characters ($m,n,...$) or an underline, $\underline{z}$, and curved
indices are written with Greek letters $\mu,\nu...$ . A hat on an index or field denotes it as an
eleven dimensional object.

For the Dirac matrices we choose the basis where $\Gamma_{\underline0\underline1\underline2...\natural}=1$ and the Majorana
conjugate is given by $\bar{\eta}=\eta^T C$, with $C$ the charge conjugation
matrix, chosen to be $C=\Gamma_{\underline{0}}$.

The action of CJS supergravity is given by \cite{Cremmer:1978km}
\ba
{\cal L}_{11}&=&\frac{1}{2\kappa^2}\left[R
        -\half\bar{\Psi}_{\hat{m}}\Gamma^{\hat{m}\hat{n}\hat{p}}\hat{D}_{\hat{n}}\Psi_{\hat{p}}
        -\half\frac{1}{4!}\hat{G}_{\hat{m}\hat{n}\hat{p}\hat{q}}G^{\hat{m}\hat{n}\hat{p}\hat{q}}\right.\\\nonumber
        &~&+\left.\frac{1}{(12)^4}\hat{\epsilon}^{\hat{m1}...\hat{n1}...\hat{p1}...}
                  \hat{G}_{\hat{m1}...}\hat{G}_{\hat{n1}...}\hat{C}_{\hat{p1}...}+...\right],
\ea
where $\hat{G}={\rm d}\hat{C}$ is the four form field strength for $\hat{C}$.
Writing the 11D vierbein as $E_{\hat{\mu}}^{\;\;\hat{m}}$ one has that
supersymmetry requires
\ba
\label{11Dsusy}
\hat{\delta}\Psi_{\hat{\mu}}=\hat{\nabla}_{\hat{\mu}}\hat{\epsilon}
              +\frac{1}{288}\left[ \hat\Gamma_{\hat{\mu}}^{\;\;\hat{m}\hat{n}\hat{p}\hat{q}}
                                  -8\delta_{\hat{\mu}}^{\hat{m}}\hat\Gamma^{\hat{n}\hat{p}\hat{q}}\right]\hat{\epsilon}
                \; G_{\hat{m}\hat{n}\hat{p}\hat{q}}=0.
\ea

To perform the dimensional reduction we write the standard triangular vielbein ansatz
leading to a metric of the form
\ba
d\hat{s}^2&=&\exp(-2\phi/3)ds^2+\exp(4\phi/3)(dz+A)^2,
\ea
which gives the action in ten dimensions in the string frame. The vielbeins are
related by \mbox{$E^a=\exp(-\phi/3)e^a$}, with $E^{\underline{z}}=\exp(2\phi/3)(dz+A)$.
Such a reduction introduces a scalar field $\phi$ and a one form $A$, with field
strength $F=dA$, into the 10D spectrum. The three-form 
and four-form field strength are then decomposed as
\ba
\hat{C}&=&C+B\wedge dz,\\
\hat{G}&=&\tilde{G}+H\wedge (dz+A),
\ea
where we have defined
\ba
H&=&dB,\\
\label{Gtilde}
\tilde{G}&=&dC-H\wedge A.
\ea
The gravitino decomposes as
\ba
\Psi_{\underline{z}}&=&\frac{1}{3}\exp(\phi/6)\Gamma_{\underline{z}}\lambda,\\\nonumber
\Psi_m&=&\exp(\phi/6)\left[\psi_m-\frac{1}{6}\Gamma_m\lambda\right],\\\nonumber
\hat{\epsilon}&=&\exp(-\phi/6)\epsilon,
\ea
which gives the following susy variations.
\ba
\label{IIAdilVar}
\delta\lambda&=&\left[\del_a\phi\Gamma^a-\frac{1}{12}H_{abc}\Gamma^{abc}\Gamma_{\underline{z}}\right]\epsilon
                    -\frac{1}{8}\exp(\phi)\left[3F_{ab}\Gamma^{ab}\Gamma_{\underline{z}}
                                               -\frac{1}{12}\tilde{G}_{abcd}\Gamma^{abcd}\right]\epsilon,\\
\label{IIAgravVar}
\delta\psi_m&=&D_m\eta-\frac{1}{8}H_{mbc}\Gamma^{bc}\Gamma_{\underline{z}}\epsilon
    -\frac{1}{8}\exp(\phi)\left[\half F_{ab}\Gamma^{ab}\Gamma_m\Gamma_{\underline{z}}
                                   -\frac{1}{4!}\tilde{G}_{abcd}\Gamma^{abcd}\Gamma_m\right]\epsilon.
\ea

\section{Some IIA p-brane solutions.}
\label{AppB}
\subsection{D0-brane}

The D0 brane solution can be derived from the M-wave by dimensional reduction
to give
\ba
ds^2_{10}&=&-U^{-\frac{1}{2}}dt^2+U^{\frac{1}{2}}d\underline{x}^2,\\\nonumber
A&=&U^{-1}(1-U)dt,\qquad\Rightarrow F=-U^{-2}dU\wedge dt,\\\nonumber
\exp(\phi)&=&U^{\frac{3}{4}},\\\nonumber
\epsilon&=&\exp(\phi/6)\hat\epsilon=U^{-\frac{1}{8}}\epsilon_0,
\ea
where $U$ is some harmonic function of $\underline{x}$ and
$\epsilon_0$ satisfies the projection $\Gamma_{\underline{0}}\Gamma_{\underline{z}}\epsilon_0=\epsilon_0$.
In order to work with a Killing spinor having a single degree of freedom we may
also choose the compatible projections,
\ba
\Gamma_{\underline{1}\underline{2}\underline{3}\underline{4}}\epsilon_0,
=\Gamma_{\underline{1}\underline{2}\underline{5}\underline{6}}\epsilon_0,
=\Gamma_{\underline{1}\underline{2}\underline{7}\underline{8}}\epsilon_0,
=\Gamma_{\underline{1}\underline{3}\underline{5}\underline{7}}\epsilon_0=\epsilon_0,
\ea
which gives us the following bilinears
\ba
X&=&U^{-\frac{1}{4}},\\
K&=&-U^{-\frac{1}{4}}e^{\underline 0},\\
\tilde{K}&=&-U^{-\frac{1}{4}}e^{\underline 9},\\
\Omega&=&-U^{-\frac{1}{4}}e^{\underline 0\underline 9},\\
Z&=&U^{-\frac{1}{4}}[e^{1234}+...],\\
\Sigma&=&-U^{-\frac{1}{4}}e^0[e^{1234}+...].
\ea

\subsection{F1-string}
\label{F1section}
Taking the M2-brane in 11 dimensions and reducing to IIA gives the following string
solution,
\ba
\label{stringSolution}
ds^2_{10}&=&U^{-1}ds^2(\bM^2)+ds^2(\bE^8),\\\nonumber
H&=&-d(U^{-1})\wedge Vol(\bM^2),\\\nonumber
B&=&-U^{-1}dt\wedge dx+dt\wedge dx,\\\nonumber
\exp(\phi)&=&U^{-\half},\\\nonumber
\epsilon&=&\exp(\phi/6)\hat\epsilon=U^{-\frac{1}{4}}\epsilon_0,
\ea
Where we have chosen a gauge for $B$ such that $B$ vanishes asymptotically,
$U$ is some harmonic function on $\bE^8$ and
$\Gamma_{\underline{0}\underline{1}\underline{z}}\epsilon_0=\epsilon_0$.
We may also make the following compatible projections so as $\epsilon$
has only one degree of freedom,
\ba\nonumber
\Gamma_{\underline{0}\underline{2}\underline{3}}\epsilon_0,
=\Gamma_{\underline{0}\underline{4}\underline{5}}\epsilon_0,
=\Gamma_{\underline{0}\underline{6}\underline{7}}\epsilon_0,
=\Gamma_{\underline{0}\underline{1}\underline{2}\underline{4}\underline{6}\underline{8}}\epsilon_0,
=\epsilon_0.
\ea
We then find that the set of bi-linears is
\ba\nonumber
X&=&0\\
K&=&-U^{-\frac{1}{2}}e^0,\\
\tilde{K}&=&U^{-\frac{1}{2}}e^1,\\
\Omega&=&U^{-\frac{1}{2}}[e^{23}+e^{45}+e^{67}+e^{89}],\\
Z&=&U^{-\frac{1}{2}}[-e^{3468}+...]+U^{-\frac{1}{2}}e^{01}[e^{23}+e^{45}+e^{67}+e^{89}],\\
\Sigma&=&U^{-\frac{1}{2}}[e^1(e^{2468}+...)+e^0(e^{2345}+...)].
\ea
\subsection{D2-brane}
\label{D2section}
Taking the smeared M2-brane in 11 dimensions and reducing to IIA gives the following brane
solution,
\ba\nonumber
ds^2_{10}&=&U^{-\frac{1}{2}}ds^2(\bM^3)+U^{\frac{1}{2}}ds^2(\bE^7),\\\nonumber
\tilde{G}&=&dC_3=-d(U^{-1})\wedge Vol(\bM^3)=U^{-\frac{5}{4}}dUe^{012},\\\nonumber
C_3&=&-U^{-1}dt\wedge dx^1 \wedge dx^2 +dt\wedge dx^1 \wedge dx^2=-U^{-\frac{1}{4}}(1-U)e^{012},\\\nonumber
\exp(\phi)&=&U^{\frac{1}{4}},\\\nonumber
\epsilon&=&\exp(\phi/6)\hat\epsilon=U^{-\frac{1}{8}}\epsilon_0,
\ea
Where we have chosen a gauge for $B$ such that $C_3$ vanishes asymptotically,
$U$ is harmonic on $\bE^7$ and
$\Gamma_{\underline{0}\underline{1}\underline{2}}\epsilon_0=\epsilon_0$.
If we also take the projections
\ba
\nonumber
\Gamma_{\underline{0}\underline{3}\underline{4}}\epsilon_0,
=\Gamma_{\underline{0}\underline{5}\underline{6}}\epsilon_0,
=\Gamma_{\underline{0}\underline{7}\underline{8}}\epsilon_0,
=\Gamma_{\underline{0}\underline{1}\underline{3}\underline{5}\underline{7}\underline{9}}\epsilon_0,
=\epsilon_0,
\ea
then we have that the spinor bi-linears become
\ba
X&=&0,\\
K&=&-U^{-\frac{1}{4}}e^0,\\
\tilde{K}&=&U^{-\frac{1}{4}}e^9,\\
\Omega&=&U^{-\frac{1}{4}}[e^{12}+e^{34}+e^{56}+e^{78}],\\
Z&=&U^{-\frac{1}{4}}[e^{1468}+...]+U^{-\frac{1}{4}}e^{0}[e^{12}+e^{34}+e^{56}+e^{78}]e^9,\\
\Sigma&=&U^{-\frac{1}{4}}[e^{13579}+...+e^0(e^{1234}+...)].
\ea
\subsection{supertube}
\label{supertubesection}
The supergravity version of the flatspace supertube discovered in \cite{Mateos:2001qs}
was found by \cite{Emparan:2001ux} to correspond to the dimensional reduction
of an M-theory solution describing the intersection of two rotating M5 branes and an
M2 brane. After dimensional reduction the solution is as follows,
\ba
\label{supertubeMetric}
ds^2_{10}&=&-U^{-1}V^{-\frac{1}{2}}(dt-\A)^2+U^{-1}V^{\frac{1}{2}}dx^2
            +V^{\frac{1}{2}}d\underline{y}^2,\\
\exp(\phi)&=&U^{-\frac{1}{2}}V^{\frac{3}{4}},\\
\tilde{G}&=&-V^{-1}e^{01}d\A,\\
dC_3&=&-Ue^{01}d(U^{-1}\A)-U^{\half}V^{-\frac{1}{4}}\A e^1d(U^{-1}\A),\\
C_3&=&-e^{01}\A,\\
H&=&-U^{-1}e^{01}dU-U^{-\frac{1}{2}}V^{-\frac{1}{4}}e^1d\A,\\
B&=&(1-U)e^{01}+U^{\frac{1}{2}}V^{-\frac{1}{4}}e^1\A,\\
A&=&V^{-1}(dt-{\cal A})-dt,\\
F&=&dA=-U^{\frac{1}{2}}V^{-\frac{7}{4}}dVe^0-V^{-1}d\A,\\
\epsilon&=&\exp(\phi/6)\hat\epsilon=U^{-\frac{1}{4}}V^{-\frac{1}{8}}\epsilon_0,\\
\Gamma_{\underline{0}\underline{z}\underline{1}}\epsilon_0&=&
\Gamma_{\underline{0}\underline{z}}\epsilon_0=\epsilon_0,
\ea
We are also free to make the following choice of $\epsilon_0$ in order to reduce the
degrees of freedom in the spinor parameter to one.
\ba
 \Gamma_{\underline{2}\underline{3}\underline{4}\underline{5}}\epsilon_0,
=\Gamma_{\underline{2}\underline{3}\underline{6}\underline{7}}\epsilon_0,
=\Gamma_{\underline{2}\underline{3}\underline{8}\underline{9}}\epsilon_0=\epsilon_0.
\ea
The bilinears are then
\ba
X&=&U^{-\frac{1}{2}}V^{-\frac{1}{4}},\\
K&=&-U^{-\frac{1}{2}}V^{-\frac{1}{4}}e^0,\\
\tilde{K}&=&-U^{-\frac{1}{2}}V^{-\frac{1}{4}}e^1,\\
\Omega&=&-U^{-\frac{1}{2}}V^{-\frac{1}{4}}e^{01},\\
Z&=&U^{-\half}V^{-\frac{1}{4}}[e^{2345}+...],\\
\Sigma&=&U^{-\frac{1}{2}}V^{-\frac{1}{4}}[-e^0(e^{2345}+...)].
\ea


\end{document}